\documentclass[review]{elsarticle}

\usepackage{lineno,hyperref}
\modulolinenumbers[5]
\usepackage{subcaption}
\usepackage{xcolor}

\begin{document}

\begin{frontmatter}

\title{Experimental observation of end-pinching in the Rayleigh breakup of an electrodynamically levitated charged droplet}

\author{Mohit Singh \corref{mycorrespondingauthor}}
\cortext[mycorrespondingauthor]{Corresponding author}
\address{Department of Chemical Engineering, Indian Institute of Technology Bombay, Mumbai, India-400076}




\begin{abstract}
The experimental time-lapse images of the breakup phenomenon of a charged droplet (diameter$\sim$100$-$300 $\mu$m) levitated in an electrodynamic (ED) balance is reported. During the breakup process, a levitated charged droplet undergoes evaporation leading to a reduction in droplet size and increase in the corresponding surface charge density. As the surface charge density reaches to a critical value, known as the Rayleigh limit, the droplet undergoes breakup by forming a jet which further ejects highly charged progeny droplets. All the successive events of the droplet breakup process such as drop deformation, breakup, and relaxation of the drop back to spherical shape after ejection of progeny droplets have been recorded using the high-speed camera at 1.3 hundred thousand frames per second. The droplet is observed to eject 3-5 progeny droplets from a jet indicating end pinching mode of breakup. The jet is then observed to relax back after ejecting 31\% of the total charge and $\sim$3\% of the total mass. A suitable theory is provided to supplement the experimental observations, and a reasonable agreement is observed. Additionally, the theory is extended for the prediction of total mass loss and the entire lifetime of a charged droplet.
\end{abstract}


\end{frontmatter}


\section{Introduction}
One often encounters the charged droplets in various atmospheric and industrial processes, for example, electrified cloud droplets\cite{mason1972bakerian}, sea spray aerosols\cite{andreas1995spray}, electrospray in the contest of ion mass spectroscopy \cite{fenn1989}, aerosol generation\cite{zilch2008charge} and inkjet printing \cite{eggers1997nonlinear, basaran2002small}. Lord Rayleigh \cite{rayleigh1882} first derived the threshold charge at which the repulsive electrostatic force equals or exceeds the capillary force and the droplet becomes unstable.

In a pioneering work, Zeleny \cite{zeleny11} experimentally observed the breakup of a liquid jet issuing from an electrified capillary using an adequately high external electric field. Macky \cite{macky31} investigated the breakup of a charged water droplet in the presence of a strong electric field for the first time. He reported that at a critical applied field, the droplet elongates and liquid filaments are drawn out from the ends due to surface instability. Photographic evidence of the droplet surface deformation, breakup, and jet formation was also reported. Further, the work of Macky \cite{macky31} was supported by Taylor \cite{taylor64} using an electro-hydrostatic theory where he predicted a conical equilibrium shape with a specific cone angle for uncharged droplets in an electric field.

However, the study of an isolated charged droplet was systematically carried out by Doyle  \cite{doyle1964}, where, a Millikan oil-drop experiment setup \cite{millikan1935} was used for droplet levitation. They observed that as the size of the droplet decreases the electric stress on the surface of the droplet increases due to inherent charge. Finally, the droplet ejects 1-10 smaller, highly charged progeny droplets along with 30\% of its total charge. Like Doyle  \cite{doyle1964}, Abbas \cite{abbas1967} reported similar results for larger sized droplet. Gomez  \cite{gomez1994} showed sub-Rayleigh (70 and 80\% of the Rayleigh limit) breakup of free-falling heptane charged droplets. Although numerous experimental studies have reported the critical limit of charge on the drop for the onset of Coulombic fission, the results of these studies show discrepancies. Duft et al. \cite{duft03} provided unambiguous experimental confirmation of the Rayleigh limit of charge. However, the fission process remains unpredictable, and currently, no study can accurately predict the charge and mass loss in the process. The reported values of charge and mass losses observed during Coulombic fissions vary from 10\% to over 70\%, and 0.1\% to 30\%, respectively \cite{doyle1964,abbas1967,schweizer1971, taflin1989,gomez1994,davis1994,widmann1997,feng2001,li2005,hunter2009}.

Theoretical analyses and numerical simulations have shown that when the charge on the drop is equal to the Rayleigh limit, an initially perturbed drop (from the spherical shape) develops conical tips at the poles, and finally, a thin filament-like jet emerges from the tips ($\sim$ ref \cite{gawande2019rayleigh,  gawande2017numerical}) which confirm the experimental photographs of Duft  \cite{duft03}. Although Duft et al., \cite{duft03} reported that the experiments are highly reproducible, the limitations of the study were the sequential images correspond to different experiments and not to the same drop in a single experiment. Very recently, Singh  \cite{singh2019effect} showed the frame-wise details of the drop deformation, breakup via jet detachment, and relaxation of drop shape after the breakup using high-speed imaging of a single droplet in a single experiment. An estimate of charge and mass loss during the breakup process is also reported. Similar to this study \cite{singh2019effect}, in the present work, we have reported high-speed imaging of a levitated charged droplet and observed an entirely distinct mode of breakup that is end-pinch off mode, where progeny droplets are ejected from the tip of the jet carrying 29 to 40\% of the original charge and $\sim$2-5\% mass. A similar mode of a jet breakup is also observed in the electrospray system by Ga{\~n}{\'a}n-Calvo  \cite{ganan2016onset}. Unlike Ga{\~n}{\'a}n-Calvo  \cite{ganan2016onset} where the first ejected progeny droplet in electrospray is examined, the present study focuses on the breakup of a levitated charged droplet.   

To the best of our knowledge, this is the first experimental study where an end pinching mode of droplet breakup and relaxation of jet after the breakup is observed in the case of a levitated charged droplet. Although the breakup of a levitated charged drop is previously reported by only two groups, namely, Duft et al., \citep{duft03} and our group  Singh et al., \citep{singh2019effect}, the dynamics of a jet, detachment or relaxation, was not clear from the images provided by Duft et al., \cite{duft03}, whereas, Singh et al., \cite{singh2019effect} have reported only jet detachment mode through high-speed imaging. Here we report the entirely different mode of breakup and jet dynamics where the droplet is observed to eject 3 to 5 progeny droplets via an end-pinching mode of the breakup, and after the ejection of progeny droplets, the jet relaxes back. Since the size of progeny droplets ejected from the drop depends on the jet characteristics and also, the mono-dispersity of the progeny droplets is highly desirable in many engineering applications such as ink-jet, fuel atomization and spray painting it is, therefore, pertinent to examine the droplet breakup and jet characteristics. It should be noted that the highlight of the manuscript is the first of its kind of experimental observation. At the same time, the back-of-the-envelope calculation is performed to support the experimental observations and to provide a few general scaling laws for the number of progeny droplets and jet diameter. Although the droplet breakup and progeny formation depend on several parameters, as shown by Ga{\~n}{\'a}n-Calvo  \cite{ganan2016onset}, the theoretical analysis in the present work is carried out based on experimentally known quantities such as the fraction of charge ($q$) and mass ($m$) loss. The magnitude of charge loss and mass loss depends on the various quantities such as surface tension, conductivity and viscosity. Thus, if one knows the exact value of $q$ and $m$ loss, the theory presented would suffice to calculate the characteristics progeny droplets thus formed. A remarkable agreement is observed between experiments and scaling laws. Additionally, the theoretical results are compared with one experimental observation, while it can be extended to other experiments also. 

\begin{figure}
	\centering
	\includegraphics[width=1\textwidth]{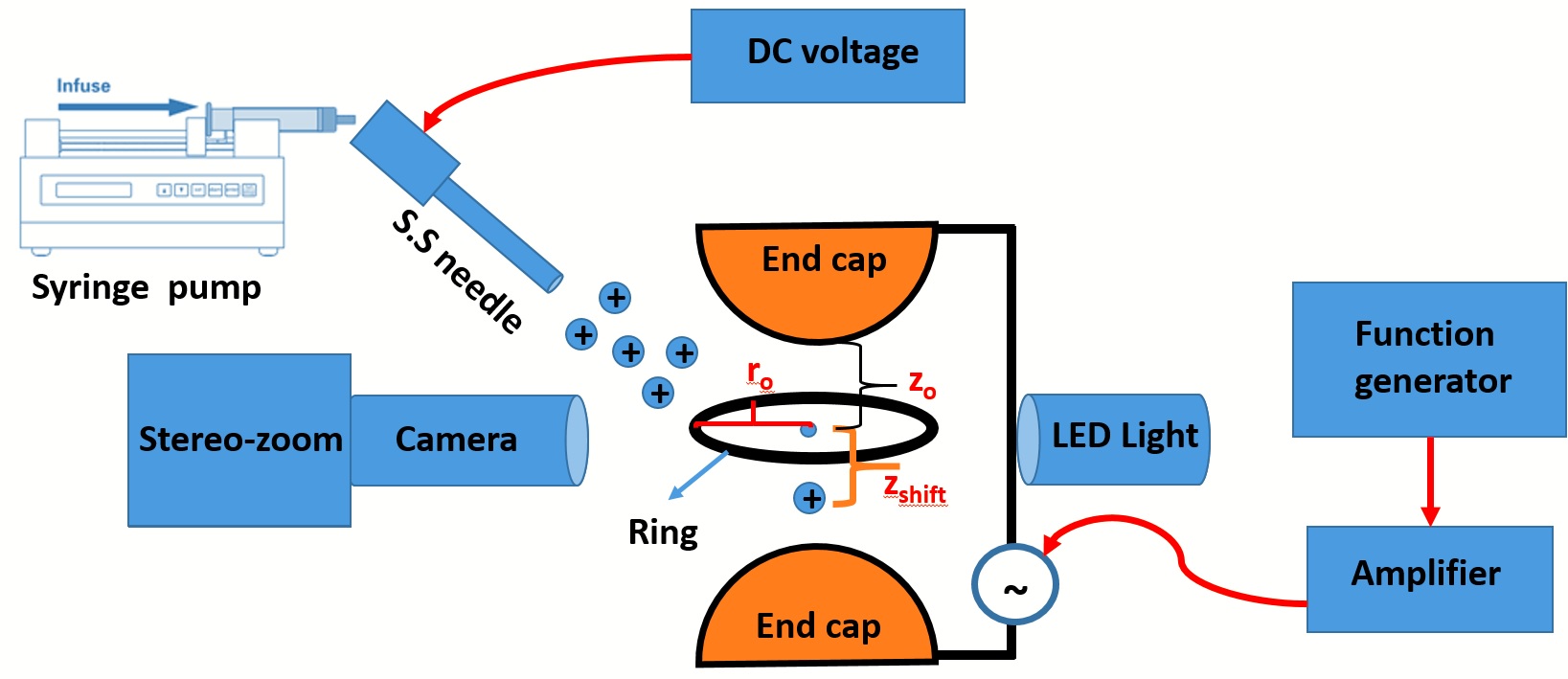}
	\caption{\textcolor{black}{Schematic of setup used for droplet levitation and charging}.}
	\label{fig:setup}
\end{figure} 
\section{Experimental} 
\subsection{Materials and method}
The experiments described in the present work involve the levitation of ethylene glycol (EG) and ethanol solution (50\% v/v) droplets. The charged droplets are generated using electrospray in the dripping mode, where a high positive DC potential (5kV) was applied on the stainless steel needle, as shown in fig.~\ref{fig:setup}. The needle was kept 40-50 mm away from the grounded electrode (the ring electrode of the electrodynamic trap). It should be noted that the electrospray is used only for the generation of the charged droplets, and the droplets are further levitated in an electrodynamic trap using quadrupolar AC potential. The appropriate amount of NaCl is added to increase the electrical conductivity of the droplet and measured using a conductivity meter (Hanna instruments, HI 2316). The viscosity of the droplet is measured using as Ostwald's viscometer and the value obtained as 0.006 Pa-s. The surface tension of the droplet is measured using the pendant drop (DIGIDROP, model DS) method and spinning drop (dataphysics, SVT 20 ) method and the values obtained as 30-40 mN/m. The experiments are carried out at normal atmospheric conditions (1 atm pressure and 25 $^\circ$C temperature).  

In the present work, a positively charged droplet is levitated in a modified Paul trap, as shown in fig.~\ref{fig:setup}. 
The trap consists of two endcap electrodes and a ring electrode. The highlight of the present trap is the higher value of $z_0$ ($\sim$6 mm, the distance between the centre of the ring and the bottom centre of the end cap electrode) and $r_0$ ($\sim$6 mm, the distance between centre of the ring and the inner periphery of the ring electrode) which provides enough space to perform several activities simultaneously such as introducing charged droplets generated by electrosprays, illuminating the drop using LED light and recording the drop deformation followed by breakup using high-speed camera (by Phantom V12 camera) at 1-1.3 hundred thousand fps. Both the endcap electrodes are shorted, and 11 $\mathrm{kV_{pp}}$ voltage is applied using a high voltage amplifier (trek 8080), which is connected with a function generator for generating the desired waveform. The ring electrode is kept grounded. The voltage was kept highest and constant to ensure the high center of mass stability of a levitated charged droplet. The imposed frequency of an AC field is varied from 100-500 Hz for stable levitation of the droplet. The camera can record upto 180 thousand fps at 128$\times$128 resolution with 2s recording time and was kept inclined at $30^0$-$40^0$ for visualization of the phenomenon. Nikon halogen light (150 W) was used as a light source to illuminate the levitated droplet.  

\subsection{charge and mass loss measurement}
Recently, Singh et al., \cite{singh2017} have reported several ways to measure the charge on the drop before and after the breakup. In the present experiments, two methods are employed to measure the charge on the droplet;
\begin{itemize}
	\item Cut-off frequency method: It is also called a destabilization method. In this method, the frequency at which a drop centre of mass motion (COM) becomes unstable was determined by gradually decreasing the applied frequency until the drop attains violent (large amplitude) COM oscillations. This frequency may be termed as the “cut-off” frequency. From the stability analysis (see ref~\cite{singh2017} ), the value of charge on the drop can be obtained. 
	\item In the second method, we compared the experimental COM oscillation dynamics of the drop with the numerical solution of the modified Mathieu (see ref~\cite{singh2019effect}, eq. 1) equation for all experimentally measured parameters except charge.     
\end{itemize}
The mass loss measurement in the breakup process is done by directly measuring the sizes of the progeny droplet (shown in fig.~\ref{fig:jet} and \ref{fig:progeny}). The image is shown in fig.~\ref{fig:jet} looks blur due to stretching and enlargement while the images are shown in fig.~\ref{fig:progeny} have sharper boundaries because they are extracted from the high-speed video at its resolution. With the help of precise image greyscale and blur thresholding using ImageJ software, one can measure the sizes accurately. The size of the progeny is measured by tracking the change of greyscale values horizontally. The distance between the sharp increase and a sharp decrease in the greyscale value is the diameter of the progeny droplet. However, we have determined the mass loss within $\pm$ 10\% of the experimental error. The 10\% error accounts for the uncertainty caused by various image corrections.

\subsection{shape fitting} \label{p3}
\begin{figure*}[t]
	\centering		
	\begin{subfigure}{0.4\linewidth}\centering
		\includegraphics[width=0.7\linewidth]{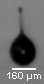}
		\caption{}
		\label{fig:jet}
	\end{subfigure}
	\begin{subfigure}{0.4\linewidth}\centering
		\includegraphics[width=1.5\linewidth]{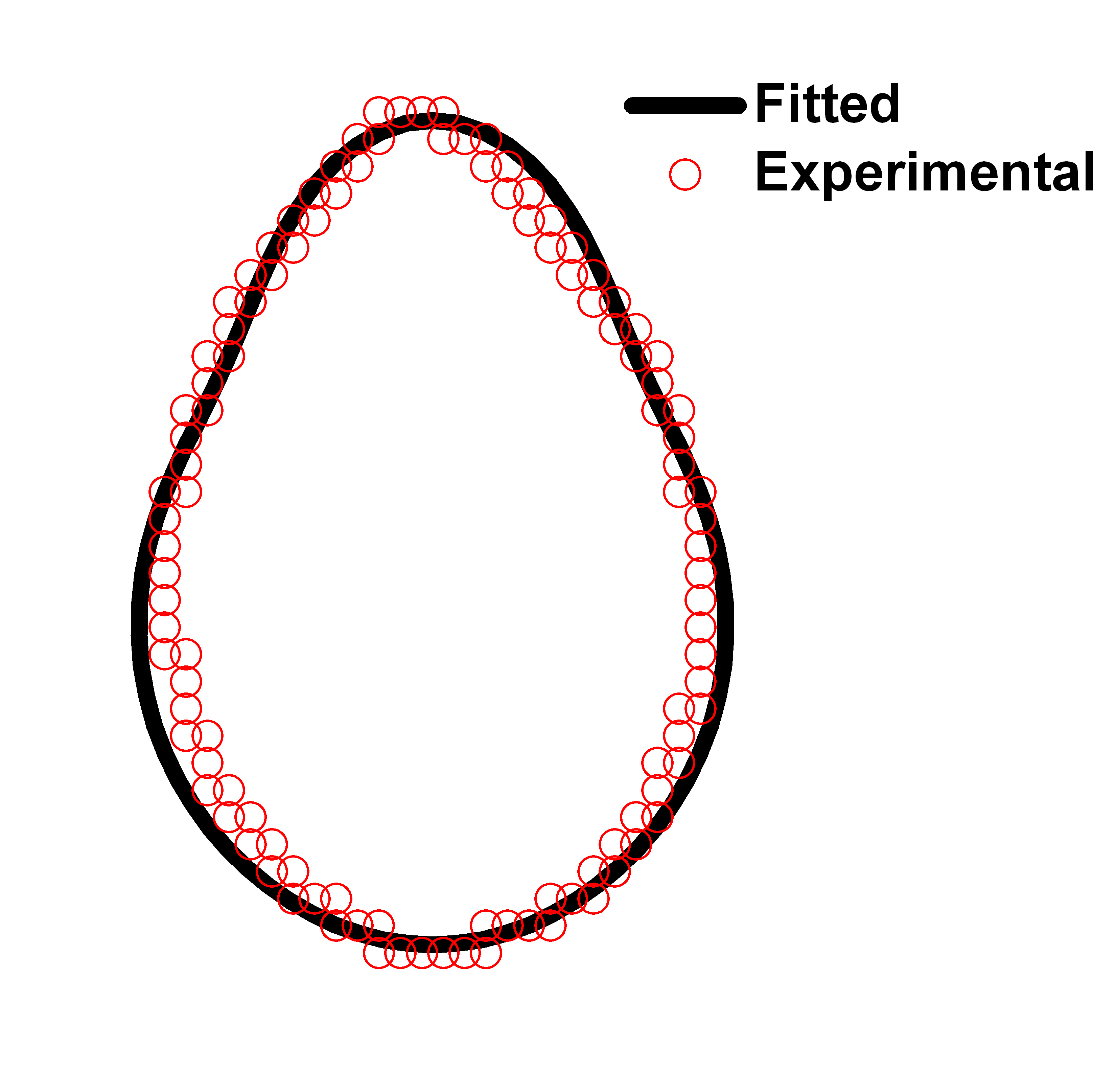}
		\caption{}
		\label{fig:shape_fit}
	\end{subfigure}
\caption{a) An enlarged image of droplet in the act of breakup. The blurriness observed in the images is due to recording at low resolution and stretching to enlarge, b) Overlapping of shape obtained from the experiments where droplet breaks in the upward direction and the shape obtained from non-linear least square fitting.}
\end{figure*}

Since an AC quadrupole field is used for levitation of a positively charged drop, the relative potential (positive or negative) of the end cap and the corresponding deformation is critical to asymmetric
breakup and the direction of the jet. In the experiments, it is observed that in most of the cases, the droplet breaks in the upward direction (positive z-direction) at the north pole. The upward breakup of a charged droplet can be explained from its shape at the starting point of the continuous deformation which corresponds to image at t=0 in figure \ref{fig:progeny}. The outline of the droplet in the image at t=0 is obtained using the ImageJ software. This outline is then fitted using the non-linear least square method to a Legendre series (using Mathematica software) to obtain the coefficients of the different Legendre modes which are responsible for the shape of the drop. The shape obtained by tracking the outline of the experimental drop and the shape obtained by fitting is plotted together and shown in figure \ref{fig:shape_fit}. It can be observed that fitted shape collapses well on the experimental shape. The equation of the shape obtained from such a fit is given below:
\begin{equation}
r_s(\theta)=R_0+\alpha_1 P_1(\cos \theta)+\alpha_2 P_2(\cos \theta) +\alpha_3 P_3(\cos \theta)+\alpha_4 P_4(\cos \theta) \label{Eq:positive_p3}
\end{equation}
The value of the coefficient of the second Legendre($P_2$) mode is 1.0, the value of third Legendre($P_3$) mode coefficient is +1.8, the value of fourth Legendre($P_4$) mode coefficient is +0.7 and $R_0=13.50$. Note that the numerical value of $R_0$ is in terms of pixels and one pixel corresponds to $\sim$ 12$\mu$m. The $P_2$ mode contributes to symmetric deformation of the droplet while the $P_3$ mode includes asymmetry in the shape of the drop. The high positive value of $P_3$ mode means a higher curvature at the north-pole while negative value indicates higher curvature at the south pole.

\subsection{various sources of error}
Since the high-speed video recording of $\sim$ 100-200 $\mu$m diameter droplet is performed at very low resolution due to the limitation of camera specification, the blurriness in the image gives error in the exact size measurement. The video was further processed using ImageJ software where image greyscale thresholding and blur thresholding may cause an error in measurement of various droplet dimensions. Thus, the standard deviation in the data accounts all these sources of error. It was observed in the various trials of the measurements that camera inclination, i.e., 30 to 40$^o$, cause 2 \% of mean error in the vertical direction measurements and, therefore diameter is measured horizontally. The measurement trials are made by imaging and measuring the known circle printed on a glass slide.

\section{Results and discussion}
\subsection*{Various experimental observations and explanations}
\begin{figure*}
	\centering
	\includegraphics[width=1\textwidth]{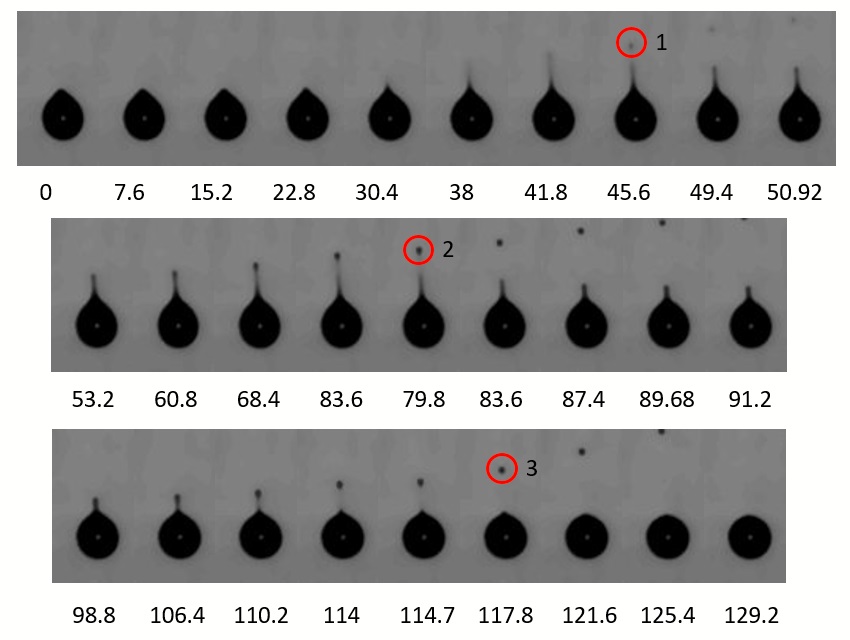}
	\caption{\textcolor{black}{Deformation, breakup, and jet and surface relaxation sequence of a levitated droplet in the process of end pinching mode of breakup. All the three ejected progenies are highlighted via red color circles numbered form 1 to 3. The numbers below the figures are time in $\mathrm{\mu}$s. Experimental parameters: applied potential ($\phi_0$)=11 $\mathrm{kV_{pp}}$, imposed frequency ($f$)=150 Hz, Droplet diameter ($D_d$)= 160 $\pm$7 $\mu$m, frame rate= 130 fps, camera resolution= 128$\times$128.}}
	\label{fig:progeny}
\end{figure*}
\begin{figure*}[htb]
	\centering
	\includegraphics[width=1\textwidth]{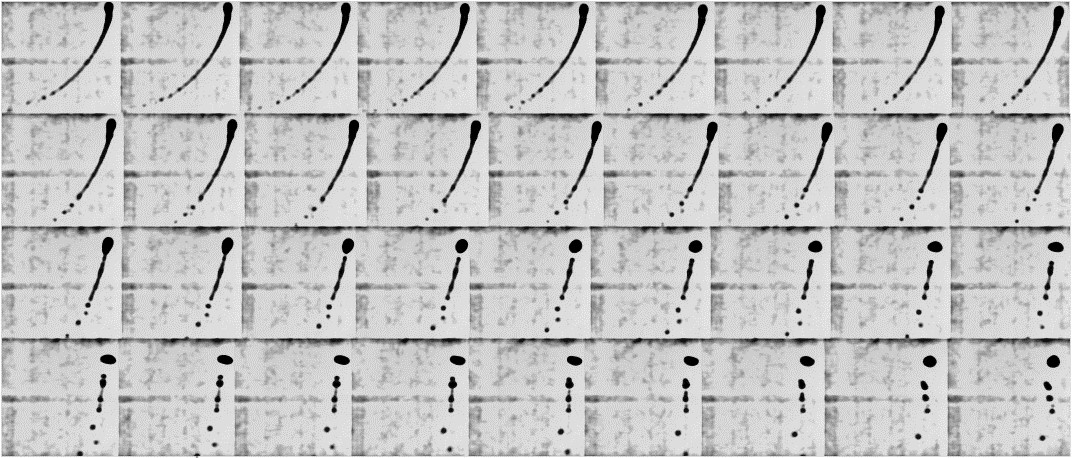}
	\caption{\textcolor{black}{High speed images of charged droplet breakup in the electro-spray setup. Parameter: Needle diameter = 50 $\mu$m, electrode spacing =80 mm, applied voltage = 7 kV, liquid= EG, type of voltage= Positive DC, model of electrospray = dripping mode, capturing location= 20 mm away from the tip of the needle, frame rate= 1.5 hundred thousand, resolution= 128 $\times$128.}}
	\label{fig:end_pinch_ES}
\end{figure*}

In a typical experiment, the levitated sub-Rayleigh charged drop undergoes evaporation and builds surface charge density with time. When the charge density exceeds beyond a critical limit, the droplet surface becomes unstable and deforms progressively to form jet and eventually breaks. The typical droplet breakup mechanism, such as drop deformation, jet formation, and breakup is shown in fig.~\ref{fig:progeny}. It can be observed from the figure that the droplet breaks in the upward direction with an up-down asymmetry. The jet is observed to break in an end pinch-off manner and relaxes back after ejecting a fraction of charge with a few countable numbers of progeny droplets. We have conducted numerous experiments by levitating different composition (varying \% volume fraction of ethanol and EG) and sized droplet at various frequencies and operating voltages. The end pinch-off mode of the droplet breakup is observed only in 20\% of the experiments. In 80\% experiments, a jet detachment mode of droplet breakup is observed where the droplet ejects a jet, and the jet further breaks into several progeny droplets, for more details of this mode of breakup refer Singh \cite{singh2019effect}. The specific circumstances for the reproducibility of the end pinch-off mode of droplet breakup could not be determined. However, a similar kind of breakup mode is observed in the electrospray setup, as shown in fig.~\ref{fig:end_pinch_ES}, and also reported in the literature ($\sim$ see ref~\cite{ganan2016onset, gomez1994}). The breakup showed in fig.~\ref{fig:end_pinch_ES}, maybe a chance encounter event, is observed in the downstream of the electrospray using a high-speed camera at 1.5 hundred thousand frames rate. It should be noted that the aim of fig.~\ref{fig:end_pinch_ES} is to show that end-pinching commonly occurs in the electrospray while the same mode of the breakup is not reported in case of isolated charged droplet breakup. One can hypothesis that the breakup observed in the electrospray is due to existence of the high field, i.e., $O(20kV/cm)$\cite{grimm2005dynamics,fontelos2008evolution}, whereas, the field applied in the levitation is $O(8kV/cm)$. Grimm \cite{grimm2005dynamics} studied the distortion, jetting, and progeny formation from charged and neutral methanol droplets subjected to a strong electric field. However, they reported only jet detachment mode of a breakup while no evidence of end-pinching. Thus, the breakup reported in the present manuscript is new and quite intriguing.    
\subsection*{Explanation of upward and asymmetric breakup}
Since the droplet is levitated in the presence of pure AC quadrupole field without superimposing any additional DC bias voltage to balance the gravity force, the droplet levitates slightly away from the geometric centre of the trap as shown in fig.~\ref{fig:setup}. At this location, the drop experiences a quadrupolarly induced uniform field ($E$=4$\Lambda$$z_\mathrm{shift}$, where, $\Lambda$ is the intensity of the quadrupole field, $z_\mathrm{shift}$ is the $z$-directional downward distance from the centre of the trap, see fig.~\ref{fig:setup}) which causes differential electrical stress at the north pole and the south pole of the drop. 

The droplet, therefore, breaks asymmetrically. In the first frame of the fig.~\ref{fig:progeny}, the droplet is observed to form a pear shape, which suggests that the drop shape has a high magnitude of 3$^\mathrm{rd}$ Legendre mode ($P_3$) perturbation, i.e., $P_3\sim$1.8. The value of the coefficient of $P_3$ is obtained by fitting the outline of the experimental drop shape with a shape equation described in terms of Legendre modes, using a nonlinear least-square fitting method (see section~\ref{p3})).  When a charged droplet continues to evaporate, the surface charge density of the drop increases and eventually reaches its Rayleigh limit. \textcolor{black}{The classical expression for the critical charge, also known as Rayleigh limit, is given by, $Q_R= 8\pi\sqrt{\epsilon \gamma R_0{^3}}$, where $\epsilon$ is the permittivity of the surrounding medium, $\gamma$ is the surface tension of the drop and $R_0$ is the droplet radius. The expression indicates that the Rayleigh limit of charge is proportional to the size of the droplet. Thus, smaller sized droplet requires a lower charge to attain its Rayleigh limit ($Q_R$). It should be noted that during droplet evaporation, the magnitude of charge remains the same, while the charge density increases due to a reduction in size.} The high charge density with high initial positive $P_3$ perturbation results in instability that causes the droplet to break in the upward direction (see ref~Singh \cite{singh2019subcritical} for detailed explanation).

\section{Theoretical prediction} 
The droplet breakup is a result of an imbalance between the destabilizing electrical stress due to charge on the drop and the stabilizing capillary stress due to the surface tension. At equilibrium, the electrical stresses which act normal to the surface of the drop equal the surface tension force. When a droplet of radius $R_0$ forms a jet during its breakup process, several progeny droplets are ejected from the tip of the jet. We have assumed the shape of the jet as a cylinder having a total surface charge $Q$. Although the shape of the drop tip is conical at the breakup, the jet can be approximated as a cylinder for the ease of theoretical analysis. Upon performing the stress balance on the cylindrical jet and using Gauss's Law the expression for the jet radius ($a$) can be obtained as,

\begin{equation}
a^3=(8 \epsilon \epsilon_0 \gamma)\frac{V^2}{Q^2}\,.
\label{4a}
\end{equation}
Since it is difficult to measure the absolute value of charge in the jet region, eq.~\ref{4a} can be translated in terms of known quantities such as charge and mass loss fractions.
 Using $V$=$f_vV_o$ and $Q$=$f_c Q_o$, where $f_v$ and $f_c$ are the lost volume and charge fractions, respectively, $V_0$ is the volume of original spherical drop and  
 $Q{_o}^2$=$48 \pi \epsilon \epsilon_0 \gamma V_0$, is the Rayleigh critical charge required for the drop to become unstable.  The most convenient expression for a radius of the cylinder, in the form of experimentally measurable parameters such as $R_o$, $f{_v}$ and $f{_c}$, reduces to,
\begin{equation}
a=R_\mathrm{jet}=R_{0}\left[\frac{2}{9} \frac{f{_v}^2}{f{_c}^2}\right]^{1/3}
\label{9}
\end{equation} 
The intermediate steps involved in the derivation are given in the supplementary file. Substituting the experimentally measured quatities such as $f_v$ ($\sim0.03$), $f_c$ ($\sim0.31$) and $R_0$ ($\sim 80 \times10^{-6}$ m) in eq.~\ref{9}, the value of jet diameter ($d_\mathrm{jet}$) can be obtained as,  
\begin{equation}
d_\mathrm{jet}=2 R_\mathrm{jet}=2\times80\times10^{-6}\left(\frac{2}{9}\frac{0.03^2}{0.31^2}\right)^{1/3}=21\ \mathrm{\mathrm{\mu m}}.
\end{equation}
Experimentally, the jet diameter is measured using ImageJ software via tracking the change in the greyscale values. The point of measurement is chosen as the intersection of two tangents drawn at the endpoint of the cone and the start of the jet. The scale used for measurement is shown in fig.~\ref{fig:jet}. The experimental value of $d_\mathrm{jet}$ is $\sim$ 23 $\mathrm{\mu}$m, which is in reasonable agreement with theoretically obtained value. Hunter and Ray \cite{hunter2009} report a similar analysis based on charge and mass conservation, but the comparison with the experimental observation of progeny droplets is not attempted in their study.

The approach can be continued to estimate the number of progeny droplet expelled during the droplet breakup process. The detailed derivation is given in the supplementary file, and the finally obtained expression for the number of progeny droplets (n) is given here as,
\begin{equation}
n=\frac{f{_c}^2}{f_v} \label{12}
\end{equation}
Thus, for $f_v$$\sim$0.03 and $f_c$$\sim$0.31 the number of progeny droplets ejected during droplet breakup is,	$n$=$\frac{0.0961}{0.03}\cong 3.$
From fig.~\ref{fig:progeny}, it can be observed that during the breakup process, exactly three progeny droplets are ejected from the endpoint of the jet before it relaxes back to the sphere. Thus, theoretical calculations and experimental observations are in a fair agreement. 


Since in the present experiments we observed primary (i.e., first breakup) breakup of a levitated charge droplet another interesting question that can be asked here is what will be the extent of ejection from the parent drop? If we assume that the droplet ejects the same number of progeny droplets in each successive breakup, it is a valuable exercise to calculate the extent of mass loss during the lifetime of a charged droplet. The droplet loses its mass in two ways first is the mass loss due to breakup of droplet and second is the mass loss due to evaporation, i.e., total mass lost is equal to mass lost due to evaporation and due to Rayleigh breakup.
\begin{figure*}[htb]
	\centering
	\includegraphics[width=0.9\textwidth]{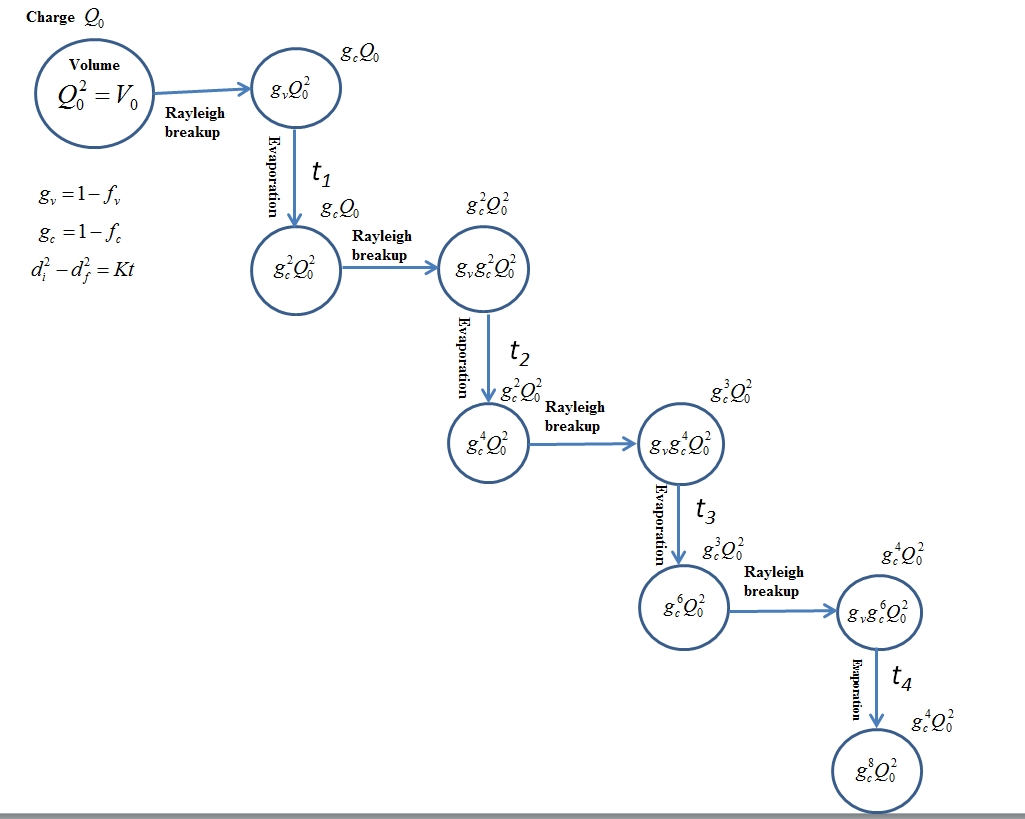}
	\caption{The flow of charge loss and mass loss of parent droplet during evaporation or breakup. Here, $f_v$ is the fraction of mass, $f_c$ is the fraction of charge loss, K is a diffusion constant, $Q_0$ is the initial charge on the droplet, $V_0$ is the initial volume of the droplet, $d_i$ and $d_f$ are initial and final diameters of drop respectively.}
	\label{fig:progeny1}
\end{figure*}

The schematic representation of the total mass loss is shown in the fig.~\ref{fig:progeny1}. The horizontal movement in the figure represents the charge and mass loss due to Rayleigh breakup, while the vertical downward movement shows the mass loss due to evaporation. The mass loss in each successive breakup is $\sim$$f_vQ_o^2$ while the remaining mass is $\sim$$(1-f_v)Q_o^2$=$g_vQ_o^2$=$g_c^2Q_o^2$. Hence, the total mass loss ($TML$) can be calculated by referring fig.~\ref{fig:progeny1}, given as,
\begin{equation}
TML =f_v+f_vg{_c}^2+f_vg{_c}^4+f_vg{_c}^6+f_vg{_c}^8 \label{14}
\end{equation}
The eq.~\ref{14} is the simple geometric progression of $g_c$. Thus, the expression for the total mass loss can be obtained as, 
\begin{equation}
TML=\frac{f_v}{1-g{_c}^2} \label{15}
\end{equation}
For, $f_v$=0.03, $g_c$=$1-f_c$=1-0.31=0.69 and $g_v$=$1-f_v$=1-0.03=0.97, the mass lost is $0.03/(1-0.69^2)$=0.057$\sim$6\%

In the droplet breakup process, one can ask another interesting question which is what will be the total life span of a charged droplet? For example, the mass loss in the droplet breakup process takes place via two different ways, Rayleigh breakup and evaporation, before it completes its lifetime. The time taken by a droplet for evaporation is longer than that of the Rayleigh breakup time. Hence first, the time taken by the evaporation is calculated, and then the correction factor for accounting the total time of breakup is multiplied. The total lifetime of the drop can be calculated by the diagrammatic model that has been developed in fig.~\ref{fig:progeny1} to evolve the generational history of a drop. The detailed derivation is skipped here for compactness of the manuscript and can be found in the supplementary file. The final expression for total time is given as,  
%
 
\begin{equation}
t=\frac{Q_0^{4/3}}{K}\frac{[g_v^{4/3}-(g_c^2)^{4/3}]}{1-(g_c^2)^{4/3}} \label{evap_time}.
\end{equation}
It can be noticed in the eq.~\ref{evap_time} that all quantities are known except $K$, which is the diffusion coefficient. By the simple theory of diffusion and neglecting temperature change, the rate of change of volume can be given by, 
\begin{equation}
V_i^{2/3}=\left[\frac{\pi}{6}\right]^{2/3}8 D_v c_sv_mt_i=K_Dt_i,
\end{equation}  
where, $k_D$=$\left[\pi/6\right]^{2/3}8 D_v c_sv_m$ is equal to the diffusion coefficient ($K$), $v_m$=$m/\rho$ =$M/(N_A\rho_L)$, $c_s$=$p_s/(k_B T)$=$p_{eq}/(k_B T)$, $p_{eq}$ is equilibrium partial pressure of droplet, $D_v$ is diffusivity of the droplet, $k_B$ is the Boltzmann constant, $T$ is the temperature, $p_s$ and $p_{eq}$ are the saturation  and equilibrium pressure respectively and $\rho_L$ is the density of the liquid. 
Thus solving eq.~\ref{evap_time} for ethylene glycol droplet  by substituting experimental parameters as, $f_v$=0.03, $f_c$=0.3, $g_v$=1-$f_v$, $g_c$=1-$f_c$, $a$=$80\times 10^{-6}$ m, $k_B$=$1.381\times10^{-23}$, $T$=298.0 K, $N_{A}$=$6.02\times10^{23}$, $D_v$=$(0.108\times10^{-4})(t/293)^{1.75}$, $M$=$62.07\times10^{-3}$ Kg, $\rho$=1113.0 $\mathrm{Kg/m^3}$, $Aa$=8.7945, $Bb$=2615.4, $Cc$=244.91, $p_{eq}$=$132\times10^{(Aa - Bb/(Cc + t - 273))}$. The time for evaporation ($t_E$) is calculated as, 
$$t_E=\frac{V_i^{2/3}}{K_D}=\frac{1.66\times10^{-8}}{2.18573\times10^{-11}}=761\ \mathrm{\mu s}.$$
It can be observed that the magnitude of evaporation time obtained is lower than that of experimental observation. This can be attributed to the parameters which are taken for the calculation of diffusivity approximately represent experimental values. Further, the time taken in the Rayleigh breakup process can be obtained from the following expression,  

$$t_\mathrm{Ray}=t_E \frac{[g_v^{4/3}-(g_c^2)^{4/3}]}{1-(g_c^2)^{4/3}}=713 \mathrm{\mu s}.$$
It is interesting to note that the total time required for evaporation $t_E$ and the total time required for Rayleigh breakup $t_\mathrm{Ray}$ is almost the same. The life-time of a charged droplet can not be validated experimentally due to following experimental limitations. 

After first ejection two things happen; 1) The droplet stability changes due to change in charge to mass ratio, 2) once the droplet gets re-stabilized (by changing imposed frequency) it continues to evaporate until it develops critical charge density for the second ejection. When droplet evaporates its size continues to decrease, and it becomes difficult to visualize such a smaller droplet and its breakup. With the present experimental zooming lenses, it is challenging to observe droplet having a size of less than 3 $\mu$m diameters. Thus we were not able to validate to the life-time of a charged droplet experimentally. Additionally, it was not possible to observe the breakup of the progeny droplet since the progeny escapes from the trap as soon as it is formed, due to its very high velocity after detachment from the jet (6-10 m/s).  As the observation of Rayleigh breakup process is itself a challenging task, we have captured primary breakup of mother droplet and provided a simplified theory based on the experimental observations. Capturing successive breakup events of a charged droplet and corresponding progeny droplets is the future scope of this work.

\section{Conclusions}
High-speed imaging of the breakup process of a charged droplet levitated in an electrodynamic balance is reported in this work. To the best of our knowledge, this is the first study which shows an experimental observation of end pinching mode of levitated charged droplet breakup. A similar mode of droplet breakup is also observed using high-speed imaging of electrospray. The experiments indicate that a levitated charged droplet ejects three equal-sized highly charged progeny droplets from the tips of the drop and the jet relaxes back after ejecting 31\% charge, and about 3\% mass of the original droplet. Unlike the previously reported study, where the levitated charged drop is observed to break via jet detachment mode \cite{duft03, singh2019effect}, a distinctly different mode of jet dynamics, i.e., end-pinching followed by jet relaxation is reported in this work. Based on the experimentally measured charge and mass loss values, a simplified theory is provided to predict the jet diameter, number of progeny droplets and the life span of a levitated charged droplet. The theoretical scaling relationships thus obtained are found in a reasonable agreement with the experimental observations. 
%

\section*{Acknowledgments}
The authors would like to acknowledge Prof. Rochish Thaokar, Prof. Y. S. Mayya and Dr. Neha Gawande for their valuable insights in this work. The project is funded by BRNS, India and experiments are performed in IIT Bombay, India.  

\section*{References}

\providecommand{\noopsort}[1]{}\providecommand{\singleletter}[1]{#1}


\begin{thebibliography}{10}
	\expandafter\ifx\csname url\endcsname\relax
	\def\url#1{\texttt{#1}}\fi
	\expandafter\ifx\csname urlprefix\endcsname\relax\def\urlprefix{URL }\fi
	\expandafter\ifx\csname href\endcsname\relax
	\def\href#1#2{#2} \def\path#1{#1}\fi
	
	\bibitem{mason1972bakerian}
	B.~J. Mason, The bakerian lecture, 1971. the physics of the thunderstorm,
	Proceedings of the Royal Society of London. A. Mathematical and Physical
	Sciences 327~(1571) (1972) 433--466.
	
	\bibitem{andreas1995spray}
	E.~L. Andreas, J.~B. Edson, E.~C. Monahan, M.~P. Rouault, S.~D. Smith, The
	spray contribution to net evaporation from the sea: A review of recent
	progress, Boundary-Layer Meteorology 72~(1-2) (1995) 3--52.
	
	\bibitem{fenn1989}
	J.~B. Fenn, M.~Mann, C.~K. Meng, S.~F. Wong, C.~M. Whitehouse, Electrospray
	ionization for mass spectrometry of large biomolecules, Science 246~(4926)
	(1989) 64--71.
	
	\bibitem{zilch2008charge}
	L.~W. Zilch, J.~T. Maze, J.~W. Smith, G.~E. Ewing, M.~F. Jarrold, Charge
	separation in the aerodynamic breakup of micrometer-sized water droplets, The
	Journal of Physical Chemistry A 112~(51) (2008) 13352--13363.
	
	\bibitem{eggers1997nonlinear}
	J.~Eggers, Nonlinear dynamics and breakup of free-surface flows, Reviews of
	modern physics 69~(3) (1997) 865.
	
	\bibitem{basaran2002small}
	O.~A. Basaran, Small-scale free surface flows with breakup: Drop formation and
	emerging applications, AIChE Journal 48~(9) (2002) 1842--1848.
	
	\bibitem{rayleigh1882}
	L.~Rayleigh, On the equilibrium of liquid conducting masses charged with
	electricity, The London, Edinburgh, and Dublin Philosophical Magazine and
	Journal of Science 14~(87) (1882) 184--186.
	
	\bibitem{zeleny11}
	J.~Zeleny, On the presence in point discharge of ions of opposite sign,
	Physical Review (Series I) 33~(1) (1911) 70.
	
	\bibitem{macky31}
	W.~Macky, Some investigations on the deformation and breaking of water drops in
	strong electric fields, Proceedings of the Royal Society of London. Series A,
	Mathematical and Physical Character (1931) 565--587.
	
	\bibitem{taylor64}
	G.~Taylor, Disintegration of water drops in an electric field, in: Proceedings
	of the Royal Society of London A: Mathematical, Physical and Engineering
	Sciences, Vol. 280, The Royal Society, 1964, pp. 383--397.
	
	\bibitem{doyle1964}
	A.~Doyle, D.~R. Moffett, B.~Vonnegut, Behavior of evaporating electrically
	charged droplets, Journal of Colloid Science 19~(2) (1964) 136--143.
	
	\bibitem{millikan1935}
	R.~A. Millikan, Electrons, protons, photons, neutrons, and cosmic rays., Tech.
	rep. (1935).
	
	\bibitem{abbas1967}
	M.~Abbas, J.~Latham, The instability of evaporating charged drops, Journal of
	Fluid Mechanics 30~(4) (1967) 663--670.
	
	\bibitem{gomez1994}
	A.~Gomez, K.~Tang, Charge and fission of droplets in electrostatic sprays,
	Physics of Fluids 6~(1) (1994) 404--414.
	
	\bibitem{duft03}
	D.~Duft, T.~Achtzehn, R.~M{\"u}ller, B.~A. Huber, T.~Leisner, Coulomb fission:
	Rayleigh jets from levitated microdroplets, Nature 421~(6919) (2003)
	128--128.
	
	\bibitem{schweizer1971}
	J.~W. Schweizer, D.~Hanson, Stability limit of charged drops, Journal of
	Colloid and Interface Science 35~(3) (1971) 417--423.
	
	\bibitem{taflin1989}
	D.~C. Taflin, T.~L. Ward, E.~J. Davis, Electrified droplet fission and the
	rayleigh limit, Langmuir 5~(2) (1989) 376--384.
	
	\bibitem{davis1994}
	E.~Davis, M.~Bridges, The rayleigh limit of charge revisited: light scattering
	from exploding droplets, Journal of aerosol science 25~(6) (1994) 1179--1199.
	
	\bibitem{widmann1997}
	J.~Widmann, C.~Aardahl, E.~Davis, Observations of non-rayleigh limit explosions
	of electrodynamically levitated microdroplets, Aerosol science and technology
	27~(5) (1997) 636--648.
	
	\bibitem{feng2001}
	X.~Feng, M.~J. Bogan, G.~R. Agnes, Coulomb fission event resolved progeny
	droplet production from isolated evaporating methanol droplets, Analytical
	chemistry 73~(18) (2001) 4499--4507.
	
	\bibitem{li2005}
	K.-Y. Li, H.~Tu, A.~K. Ray, Charge limits on droplets during evaporation,
	Langmuir 21~(9) (2005) 3786--3794.
	
	\bibitem{hunter2009}
	H.~C. Hunter, A.~K. Ray, On progeny droplets emitted during coulombic fission
	of charged microdrops, Physical Chemistry Chemical Physics 11~(29) (2009)
	6156--6165.
	
	\bibitem{gawande2019rayleigh}
	N.~Gawande, Y.~Mayya, R.~Thaokar, Rayleigh breakup of a charged viscous drop
	via tip-streaming, arXiv preprint arXiv:1902.08499.
	
	\bibitem{gawande2017numerical}
	N.~Gawande, Y.~Mayya, R.~Thaokar, Numerical study of rayleigh fission of a
	charged viscous liquid drop, Physical Review Fluids 2~(11) (2017) 113603.
	
	\bibitem{singh2019effect}
	M.~Singh, N.~Gawande, Y.~Mayya, R.~Thaokar, Effect of the quadrupolar trap
	potential on the rayleigh instability and breakup of a levitated charged
	droplet, Langmuir 35~(48) (2019) 15759--15768.
	
	\bibitem{ganan2016onset}
	A.~M. Ga{\~n}{\'a}n-Calvo, J.~M. L{\'o}pez-Herrera, N.~Rebollo-Mu{\~n}oz,
	J.~Montanero, The onset of electrospray: the universal scaling laws of the
	first ejection, Scientific reports 6 (2016) 32357.
	
	\bibitem{singh2017}
	M.~Singh, Y.~Mayya, J.~Gaware, R.~M. Thaokar, Levitation dynamics of a
	collection of charged droplets in an electrodynamic balance, Journal of
	Applied Physics 121~(5) (2017) 054503.
	
	\bibitem{singh2019subcritical}
	M.~Singh, N.~Gawande, Y.~Mayya, R.~Thaokar, Subcritical asymmetric rayleigh
	breakup of a charged drop in an ac quadrupole trap, arXiv preprint
	arXiv:1907.02294.
	
	\bibitem{grimm2005dynamics}
	R.~L. Grimm, J.~L. Beauchamp, Dynamics of field-induced droplet ionization:
	Time-resolved studies of distortion, jetting, and progeny formation from
	charged and neutral methanol droplets exposed to strong electric fields, The
	Journal of Physical Chemistry B 109~(16) (2005) 8244--8250.
	
	\bibitem{fontelos2008evolution}
	M.~A. Fontelos, U.~Kindel{\'a}n, O.~Vantzos, Evolution of neutral and charged
	droplets in an electric field, Physics of Fluids 20~(9) (2008) 092110.
	
\end{thebibliography}
\end{document}